\documentclass[amsmath,amssymb,twocolumn,twoside,floatfix,nofootinbib]{revtex4}
\usepackage{natbib}
\usepackage{bbold}
\usepackage{graphicx}
\usepackage{amsfonts}
\usepackage{mathrsfs}
\usepackage{amsthm}
\usepackage[all]{xy}
\usepackage[usenames,dvipsnames]{color}
\usepackage{hyperref}
\newcommand{\beq}{\begin{equation}}
\newcommand{\eeq}{\end{equation}}
\newcommand{\bea}{\begin{eqnarray}}
\newcommand{\eea}{\end{eqnarray}}

\newcommand{\ea}{\end{align*}}

\newcommand{\bma}{\begin{pmatrix}}
\newcommand{\ema}{\end{pmatrix}}

\usepackage[percent]{overpic}
\usepackage{overpic}

\begin{document}
\title{Equiaffine braneworld}
\author{Fan Zhang} 
\affiliation{Gravitational Wave and Cosmology Laboratory, Department of Astronomy, Beijing Normal University, Beijing 100875, China}

\date{\today}

\begin{abstract}
Higher dimensional theories, wherein our four dimensional universe is immersed into a bulk ambient, have received much attention recently, and the direction of investigation had, as far as we can discern, all followed the ordinary Euclidean hypersurface theory's isometric immersion recipe, with the spacetime metric being induced by an ambient parent. We note in this paper that the indefinite signature of the Lorentzian metric perhaps hints at the lesser known equiaffine hypersurface theory as being a possibly more natural, i.e., less customized beyond minimal mathematical formalism, description of our universe's extrinsic geometry. In this alternative, the ambient is deprived of a metric, and the spacetime metric becomes conformal to the second fundamental form of the ordinary theory, therefore is automatically indefinite for hyperbolic shapes. Herein, we advocate investigations in this direction by identifying some potential physical benefits to enlisting the help of equiaffine differential geometry.
  \end{abstract}
\maketitle

\section{Introduction}
Within the mathematical literature, there are a number of different theories that were constructed to describe the extrinsic geometry of a hypersurface\footnote{Codimension one is the simplest case and appears to be sufficient for yielding the desired physical implications discussed in this paper, we note nonetheless that generalizations to higher codimensions is possible. We also assume flat ambient throughout this paper. Indeed, the integrability conditions we consider later are equivalent to demanding that the ambient curvature vanishes. However, generalizations to ambients of constant sectional curvatures should also be possible.} immersed into a bulk ambient. While the intrinsic geometry of our spacetime is verifiable best described using the pseudo-Riemannian geometry, we still have (the freedom) to choose its extrinsic counterpart, should we envision a braneworld scenario and thus a bulk ambient that physically exists. 
The various extrinsic theories differ mainly on how much machinery is made available to us. More infrastructure would facilitate more specialization, resulting in simpler expressions and easier equations to solve. Specifically, when the ambient is equipped with a metric, we can define angles and norms for the vector bases in reference frames, thus concentrate only on orthonormal frames and work with smaller principal bundles. 

Our natural bias towards convenience, coupled with the fact that the intrinsic side General Relativity is a metric theory, perhaps led to the universal assumption that the ambient should also be equipped with a metric, and that our spacetime is isometrically immersed into it. There is however the complication that the intrinsic metric is of an indefinite Lorentzian signature. Since a positive-definite ambient metric would not be able to induce a Lorentzian one on the brane, 
the ambient metric has to be indefinite as well. 
Such pseudo-metrics are rather pathological. Besides the general counter-intuitiveness of the hyperbolic trigonometry \cite{10.2307/2323737,1921Natur.108..434K}, the fact that their metric balls are noncompact complicates the search for compatible topologies into a rather messy business (see e.g., \cite{Hoekzema2011OnTT}). Such problems render the pseudo-metrics arguably quite awkward to use. 

In other words, if Nature had intended to grant us the convenience of an ambient metric, it is only doing so half-heartedly. It is then, at the very least prudent, to also consider the possibility that there is no such intention, and our intrinsic Lorentzian metric is not induced by an ambient metric, but instead bespeaks a different concept in some non-metric extrinsic theory. Such theories do exist, and one is called the equiaffine\footnote{\label{fn:naming}Or special affine or unimodular or sometimes simply affine; we will substitute out these alternative terms from literature to avoid cluttering nomenclature. Also, in deference to more established convention, we will refer to theories equipped with an ambient metric as being ``ordinary Euclidean'', even though we allow indefinite ambient metrics in the context of our physical discussion.} differential geometry, wherein a symmetric bilinear form called equiaffine metric is laid onto the immersed hypersurface, but which is in fact conformal to the second fundamental form of the immersion, in the more familiar ordinary Euclidean terminology. 

In this short note, we outline this proposal that our relativistic Lorentzian metric might profitably be interpreted as an equiaffine metric (the relevant concepts are introduced in Sec.~\ref{sec:Concepts}). Specifically, we argue in Sec.~\ref{sec:CC}, that the Einstein's equations can possibly then be regarded as the equiaffine Gauss equations in disguise, with the dark energy arising naturally through a mean curvature term. We also speculate on some potential global cosmological implications of the equiaffine braneworld scenario in Sec.~\ref{sec:Conclusion}.

\section{Concepts} \label{sec:Concepts}
Roughly speaking, equiaffine differential geometry considers concepts that are invariant under the special linear transformations and translations, which unfortunately do not include orthogonality as determined by an ambient metric. Therefore, the equiaffine differential geometry has to be constructed without summoning help from any ambient metric (thus the name ``affine''). Such more rudimentary mathematics, with less circumstance-specific appendages, are particularly likely to infiltrate physics, since they facilitate more useful physical laws that are applicable to more experimental settings and engineering situations\footnote{This is a physical representation of the Erlanger programme's rationale for developing non-Euclidean geometries. By loosening the restrictions on the transformations, so less of the geometric constructs stay invariant, we isolate and concentrate on only those properties that are the most robust, the computations concerning whom simultaneously apply to larger collections of geometric objects.}. In other words, the bottom-up experiments-driven organic growth of physical models may go against the top-down theory builders' wishes (and perhaps at times overly optimistic assumptions) to have more tools in the box.   

An immediate consequence of the depletion of the toolbox is that we cannot use the usual procedure to obtain the first fundamental form. Fortunately though, we notice that the second fundamental form $\mathbb{\Pi}_{ab}$ can still be defined and even be appropriately scaled to become equiaffine invariant \cite{BLA1}. It can thus possibly reincarnate into some sort of replacement intrinsic metric. Heuristically, since the eigenvectors of $\mathbb{\Pi}_{ab}$ are conventionally regarded as being mutually orthogonal, and the eigenvalues can be used as reference scales, the principal structure of $\mathbb{\Pi}_{ab}$ contains the necessary ingredients for the construction of an intrinsic metric gravity theory. But there is another problem. To obtain a second fundamental form, one has to have a normal first, but we cannot define a normal in the usual way by demanding it be orthogonal to the tangents of the hypersurface. An alternative procedure has to be developed, which we briefly summarize in Appendix \ref{ap:Equiaffine} (see e.g.,\cite{AffineGeo} for further details). 

Those steps are sufficient to pick out a unique bilinear form $\mathbb{\Pi}_{ab}$ associated with a unique equiaffine normal $\nu^{\alpha}$, which is then used as a metric called the equiaffine first fundamental form \cite{Spivak} or equiaffine metric \cite{Nomizu1987}, but obviously not necessarily positive-definite. Despite this equiaffine metric not being a measure of distances in the traditional sense, the equiaffine theory nevertheless, by construction and in particular due to the efforts of Blaschke and colleagues, enjoys substantial computational similarity with its Euclidean counterpart (e.g., the conditions in Appendix \ref{ap:Equiaffine} are designed to mimic properties seen in the Euclidean theory). This lack of operational distinction might have contributed to the conspicuous lack of explicit references to equiaffine geometry from the physical studies, whereby the theories are reverse-engineered to, at least initially, be phenomenological descriptions of experiments, and thus one has direct access only to the mechanics, and not the underlying geometric significances that might have set the two differential geometries apart. A potential opportunity for equiaffine geometry to break away from anonymity though, arises when both intrinsic and extrinsic curvatures make simultaneous appearances, as in the Gauss equations \eqref{eq:Gauss} below, to make hiding the double life of the equiaffine metric untenable, outing it as a curvature in actuality. We will exploit the consequences of this revelation in Sec.~\ref{sec:CC}, to model the dark energy or cosmological constant. We highlight this particular facet of cosmology, precisely because it offers a handle for us to assess the potential relevance of equiaffine geometry in a physical context. 

Besides the equiaffine metric, there is another important quantity in equiaffine geometry, the totally symmetric cubic form $\mathbb{F}_{abc}$ (see Eq.~\ref{eq:CubicFormDef} in particular), which is also called the equiaffine second fundamental form \cite{Spivak}. Just as $\mathbb{\Pi}_{ab}$ describes the next order warping of the tangential or osculating hyperplane that produces a better approximating osculating quadric, $\mathbb{F}_{abc}$ describes how the quadric can be crafted further into an even more snug-fitting osculating cubic ($\mathbb{F}_{abc}$ is related to the third order Taylor expansion coefficients in the local graph representation of the hypersurface, and $\mathbb{\Pi}_{ab}$ to the Hessian \cite{Spivak,Flanders1965}). In other words, concepts in the equiaffine theory are shifted up one order (or more, see Footnote \ref{fn:FirstCodazzi} below) as compared to the Euclidean theory.

This hike in orders implies in particular, that in the equiaffine context, the local Minkowski approximation to a spacetime is not the flat osculating hyperplane, as would be the case with isometric immersion. It is instead the osculating quadric, with (equiaffine) ``flatness'' now meaning that the third order adjustments $\mathbb{F}_{abc}$ vanish. It is more specifically a paraboloid or improper affine hypersphere to boot, meaning its equiaffine normals are parallel to each other, and it is thus also ``flat'' in the sense that its shape operator (representing fourth order adjustments, see Footnote \ref{fn:FirstCodazzi} below), given by Eq.~\ref{eq:Shape}, vanishes. Such simplicity that equiaffine geometry endows upon the Minkowski space is one of the reasons we favour equiaffine over its centroaffine or Euclidean-affine (see Appendix \ref{ap:Equiaffine}) siblings within the family of alternative affine geometries. Because we believe that when physical laws were drafted, the authors tried to, consciously or not, save as much ink as possible when it comes to describing the empty stage, and thus we enjoy better odds at finding General Relativity in the equiaffine corner of the overarching general affine ballpark. 

\section{Equations}\label{sec:CC}
The Gauss formula and Weingarten equation (together constitute the structure equations) utilized in Appendix \ref{ap:Equiaffine} break ambient derivatives of tangential and normal vector fields, respectively, into tangential and normal parts, and inform us about the bending of the hypersurface through the cross-mixings, quantified by $\mathbb{\Pi}_{ab}$ and the shape operator $\mathbb{S}_a{}^{b}$. Given an arbitrary prescription of $\mathbb{\Pi}_{ab}$ and $\mathbb{S}_a{}^{b}$ though, there is no automatic guarantee that the desired warping is achievable inside a flat ambient that is only one dimension higher. 

Taking cues from the ordinary Euclidean theory, one needs to take further derivatives, and obtain the Gauss and Codazzi equations as the integrability conditions for the structure equations. The existence part of the fundamental theorem for hypersurfaces then ensures that given a first and a second fundamental form that are compliant with these conditions, we are guaranteed a successful immersion. With the isometric immersion of spacetime in that theory though, the physical equations of motion have nothing to do with the Gauss and Codazzi equations, thus when General Relativity provides the intrinsic metric, no integrability is guaranteed \emph{a priori}, and one has yet to ensure that there exists a second fundamental form, that solves both the Gauss and Codazzi equations. In other words, we still need to find a way to warp the hypersurface, in such a way that the resulting stretching and squeezing reproduces the prescribed intrinsic distance changes. This is a highly nontrivial task (see e.g., \cite{Kasner:1921zz,Eddington1924,Deser:1976qt,Embedding2,RevModPhys.5.62,Sengor:2013jh,MR1506434,MR1506435,Fronsdal:1959zza,CMT,Paston:2013mfa}), requiring for general spacetimes a ten dimensional flat pseudo-Riemannian ambient. 

In the equiaffine theory, the fundamental forms are replaced by the equiaffine metric $\mathbb{\Pi}_{ab}$ and the cubic form $\mathbb{F}_{abc}$ (in place of $\mathbb{S}_{a}{}^{b}$, see Footnote \ref{fn:FirstCodazzi}), stipulated to satisfy a similarly named set of integrability equations \cite{AffineGeo,BLA1,DNV}. But when it comes to braneworld physics, instead of simply transcribing the Euclidean case, we make the observation that if the ambient does physically exist, then the integrability conditions must somehow already be asserting themselves, in disguise, as physical laws. Or else our theories would be yielding copious illegitimate solutions that cannot be realized in Nature because they prevent our brane from fitting into the ambient, which doesn't appear to be the situation. Specifically, we conjecture to identify Einstein's equations with the equiaffine Gauss equations, and link up matter contents with the extrinsic curvature, so that their presence warps spacetime, in a quite literal extrinsic geometric sense. This way, the metric and matter fields as solutions to the Einstein's equations and matter equations of motion serve up $\mathbb{\Pi}_{ab}$ and $\mathbb{F}_{abc}$ that are, by construction, best positioned\footnote{We caution that, by a naive counting of indices, the equiaffine integrability equations are overdetermined, so in principal, some pathological solutions we obtain using a subset of the equations (e.g., the intrinsic metric can be obtained using the Ricci part of the equiaffine Gauss equations, plus gauge conditions, without referring to the Weyl half) could possibly violate the remaining ones. The internal consistency between these equations may prevent this, as in the case of electromagnetism and General Relativity (Einstein's equations and Bianchi identities) -- after a $3+1$ split, some equations may become constraints that are preserved by the evolution equations and thus trivialize. However, we haven't managed to find theorems clarifying the extent to which this is guaranteed for the equiaffine integrability conditions, thus the circumspect statement.} to satisfy the integrability conditions\footnote{Note that while the integrability conditions take care of local immersibility, there may be additional global embeddability conditions that prevent self-intersections. They could possibly manifest as integral versions of energy conditions, with the mostly attractive nature of gravity being the consequence of it being much easier to embed an overall (local infringements can be compensated by nearby regions) more (as compared to the ambient) positively curved hypersurface (cf.~\cite{Wolfgang} vs.~\cite{Hilbert}) -- it is difficult to fit a stretchy hypersurface into a comparatively more crumpled up ambient without having to fold it back on itself.} (note in particular, the Minkowski setup of constant $\mathbb{\Pi}_{ab} = \text{diag}(-1,1,1,1)$ and $\mathbb{F}_{abc}=0$ satisfies these conditions trivially). 

Besides immersibility, there are other advantages to identifying Einstein's equations with the equiaffine Gauss equations, whose once-traced form for a four dimensional hypersurface can be turned into 
\bea \label{eq:Gauss}
{}_{\mathbb{\Pi}}R_{ab} - 3\mathcal{H} \mathbb{\Pi}_{ab} = \mathbb{F}_{ac}{}^{d}\mathbb{F}_{bd}{}^{c}-\frac{1}{2}\mathbb{F}_{ab}{}^{c}{}_{;c} \,,
\eea
after substituting in the once-traced first equiaffine Codazzi equations\footnote{\label{fn:FirstCodazzi}The first equiaffine Codazzi equation can be seen as a constraint equation relating the Weingarten form (equals shape operator with one index lowered by $\mathbb{\Pi}_{ab}$, according to the Ricci equation) $\mathbb{S}_{ab}$ and the first derivatives of $\mathbb{F}_{abc}$. In this sense, $\mathbb{S}_{ab}$ can be seen as an auxiliary variable defined to break a second order equation for $\mathbb{F}_{abc}$ down into two sets of first order equiaffine Codazzi equations.}
\bea \label{eq:FirstCodazzi}
\mathbb{S}_{ab} = \mathcal{H}\mathbb{\Pi}_{ab} - \frac{1}{2} \mathbb{F}_{ab}{}^{c}{}_{;c}\,,
\eea
where $\mathcal{H} \equiv \mathbb{S}^a{}_{a}/4$ is the equiaffine mean curvature, and semi-column denotes covariant derivative using the Levi-Civita connection of $\mathbb{\Pi}_{ab}$. Note also that the apolarity condition \eqref{eq:apolarity} enforces that the divergence term in Eq.~\eqref{eq:FirstCodazzi} is traceless. 
Using the equiaffine Theorema Egregium (obtained by taking another contraction over Eq.~\ref{eq:Gauss}), this expression can be processed further into one containing the Einstein tensor (sign conventions match those of \cite{MTW})
\begin{align} \label{eq:Einstein}
{}_{\mathbb{\Pi}}G_{ab} + 3\mathcal{H}\mathbb{\Pi}_{ab} = \mathbb{F}_{ac}{}^{d}\mathbb{F}_{bd}{}^{c}-\frac{1}{2}\mathbb{F}_{ab}{}^{c}{}_{;c} - \frac{\mathbb{F}_{def}\mathbb{F}^{def}}{2} \mathbb{\Pi}_{ab} \,.
\end{align}

In principal, $\mathcal{H}$ is a function of $\mathbb{\Pi}_{ab}$ and $\mathbb{F}_{abc}$, but it has a special status for physical membranes that have surface tension in them. Being the divergence of the equiaffine\footnote{The ensuing qualitative argument applies to either the equiaffine or the Euclidean normal, leading to similarities, e.g., within equiaffine geometry, a minimal hypersurface is still characterized by $\mathcal{H}=0$, and hyperspheres (whose equiaffine normal directions meet at a single ambient point, which can be at infinity) are still of constant $\mathcal{H}$.} normal (see Eq.~\ref{eq:Shape}), $\mathcal{H}$ measures how buckled our hypersurface is, and if it varies rapidly, we would end up with a very bumpy surface. The local high frequency bumpiness contributes nothing towards satisfying the global constraints of any variational isoperimetric problem (cf., the fixed boundary of a Plateau problem for a soap bubble supported by a wire frame, or the fixed enclosed volume for a free-floating bubble), but increases the surface area. So whatever constraints our universe has to conform to\footnote{Global constraints often restrict admissible topology, and scalars constructed out of $\mathbb{S}^a{}_b$ (i.e., functions of the coefficients in its characteristic polynomial, including $\mathcal{H}$) provide densities that integrate into topological invariants, so one would expect $\mathcal{H}$ to be inversely related to the overall sizes of the universe. Indeed, as noted by e.g.~\cite{Barrow}, inverse squareroot of the cosmological constant is on the order of $10$Gly, roughly matching the age of the universe in natural units, which is to be expected if we are not fine-tuned to reside in a special era in the history of the universe, so our distance to landmark events like the big bang is generic.}, it is quite likely that variations in $\mathcal{H}$ is suppressed (cf.~the smoothing properties of mean curvature and in particular, surface tension flows), allowing it to masquerade as the cosmological constant.
Such an entry, proportional to $\mathbb{\Pi}_{ab}$, takes an appearance in Eq.~\eqref{eq:Einstein} because $\mathbb{\Pi}_{ab}$ really describes the extrinsic shape of the hypersurface. A mean curvature rescaled second fundamental form is in fact also present in the ordinary Euclidean theory's version of the Gauss equations, but since over there, it is the first fundamental form that serves as the intrinsic metric instead, such a term cannot then readily be identified with the cosmological constant contribution.

Finally, the matter stress-energy tensor must correspond to the $\mathbb{F}_{abc}$ terms on the right hand side of Eq.~\eqref{eq:Einstein}. The second equiaffine Codazzi equations, governing the second derivatives of $\mathbb{F}_{abc}$, should then be compatible with their equations of motion. As a toy model for exploring possibilities of how this might happen, consider a situation where our region of interest, near some type of particles, happens to be approximated by a neighbourhood of an affine hypersphere (these are quite diverse and can become rather complicated, see e.g., \cite{Martinez} for some visual illustrations), then we have $\mathbb{S}_{ab} = \mathcal{H} \mathbb{\Pi}_{ab}$, or equivalently 
\bea \label{eq:Sphere}
\mathbb{F}_{ab}{}^{c}{}_{;c}=0\,, 
\eea 
according to Eq.~\eqref{eq:FirstCodazzi}, with $\mathcal{H}$ being a constant. It is easy to check that the second equiaffine Codazzi equations 
\bea \label{eq:SecondCodazzi}
\mathbb{S}_{ab;c}-\mathbb{S}_{ac;b} = \mathbb{F}_{ab}{}^d\,\mathbb{S}_{cd} - \mathbb{F}_{ac}{}^d\,\mathbb{S}_{bd} 
\eea
are satisfied (both sides vanish). In other words, Eq.~\eqref{eq:Sphere}, together with the constant $\mathcal{H}$ condition, can be regarded as a sufficient replacement to Eq.~\eqref{eq:SecondCodazzi}. Furthermore, Eq.~\eqref{eq:Sphere} also takes out the middle term on the right hand side of Eq.~\eqref{eq:Einstein}.  
The resulting expression and Eq.~\eqref{eq:Sphere} bear cursory resemblances in form to the stress-energy tensor and the sourceless equations of motion for the Yang-Mills fields, so a formal mapping fitting (some sectors of) the Standard Model into $\mathbb{F}_{abc}$ might not be prohibitively difficult to contrive. However, it would be much more satisfying if we could unearth the underlying equiaffine extrinsic geometric significances, if there is indeed any, of the Standard Model entities, an arduous task that we will relegate to future studies. We highlight here one particular difficulty, that index symmetry already restricts the number of independent degrees of freedom in $\mathbb{F}_{abc}$ to at most $20$, yet there are $118$ in the Standard Model. So unless we augment the equiaffine freedoms by e.g., increasing the codimensions, the aforementioned mapping would be highly non-injective (multiple particle configurations warp spacetime the same way) and thus non-invertible, so one perhaps shouldn't hope to recover all necessary insights by examining the equiaffine rendition of gravity alone.

\section{Discussion and Conclusion} \label{sec:Conclusion}
So far, we have confined ourselves to more local considerations. On the other hand, the equiaffine braneworld scenario has obvious global cosmological implications. For example, it is difficult to embed a compact hypersurface that is everywhere non-convex \cite{Spivak} (hard to constrict the hypersurface while not being able to find any support planes), so it is not unreasonable to consider situations where some parts of our universe are strongly convex, i.e., where the equiaffine metric is positive/negative-definite, because e.g., as a part of the isoperimetric consideration, our brane universe may need to enclose a fixed ambient volume. Any smooth change in metric signature will then inevitably lead us to a transition boundary, that can be identified with the big bang, where at least one of the eigenvalues of $\mathbb{\Pi}_{ab}$ vanishes, so it becomes degenerate, and the machinery of equiaffine geometry breaks down (at the very least $\mathbb{\Pi}^{ab}$ diverges). In other words, the big bang appears singular only because of the limitations of the particular mathematical infrastructure we implicitly adopt. Such is a rather desirable situation, since there is then no fundamental obstruction preventing us from improving our modelling and be able to impose initial conditions for our Lorentzian side of the universe. In particular, there is no need to censor the big bang, justifying excluding it from the cosmic censorship conjectures. In fact, even without major remodelling efforts, such degenerate locales can often be handled to some extent with finesse, e.g., the Frenet frame can be generalized to smoothly extend across an inflection point, where the normal of a curve is not defined \cite{CartanBook}. Similarly in the spacetime case, a degenerate boundary may be handled via the matching of limits of regular quantities obtained on either side (see e.g., \cite{1993CQGra..10.2363K,Zhang:2019rrc}).

In fact, in the ordinary Euclidean sense, the big bang would be flatter than elsewhere (cf.~the low initial gravitational entropy issues \cite{2018FoPh...48.1177P}), in the sense that $\mathbb{\Pi}_{ab}$ is conformal to the second fundamental form of the ordinary Euclidean theory, thus at least one of the Euclidean principal curvatures vanishes there. It being a highly warped place is instead in the equiaffine sense, meaning it is quite far from being a quadric, viz.~some components of $\mathbb{F}_{abc}$ are large. This is necessary for the metric signature switch to happen, because the defining quadratic polynomial $\varphi$ for which any quadric is the zero set of, would have its Hessian being conformal to $\mathbb{\Pi}_{ab}$ (see Appendix \ref{ap:Equiaffine}, and Ref.~\cite{Spivak} Vol.~3, Chap.~3 for a 2-D illustration). The Hessian of a quadratic polynomial is constant, so the big bang region would have to shun all quadric shapes in order to host a convexity change. Alternatively stated, the ambient-induced connection $\tilde{\nabla}_a$ has difficulty preserving $\mathbb{\Pi}_{bc}$ via parallel transportation when even its qualitative fundamentals like rank and signature change, thus $\mathbb{F}_{abc}$ is large according to Eq.~\ref{eq:FDown}. 

A byproduct of this feature is that the Pick invariant $\mathcal{J}\equiv \mathbb{F}_{abc}\mathbb{F}^{abc}/12$ in Eq.~\eqref{eq:Einstein} would likely end up being appreciable near the big bang, plausibly demanding the presence of a significant inflaton potential style component in the matter stress-energy tensor. In other words, inflation and the big bang come hand in hand, unless chance cancellations occur during index contractions to suppress $\mathcal{J}$ even while $\mathbb{F}_{abc}$ is component-wise (as measured by e.g., the $\text{L}_2$ or $\text{L}_{\infty}$ norm) not small. This can only happen when $\mathbb{\Pi}_{ab}$ has an indefinite signature though, and we could simply approach the big bang from the convex side and evoke continuity arguments instead, which incidentally reveals that $\mathcal{J}$ should get quite large on that side of the big bang already, and our Lorentzian side of the universe would be born directly into ongoing inflation, without any potentially problematic \cite{1992PhR...214..223G} delay. 

Due to the foundational role the Lorentzian metric plays in modern physics, there would be much to explore in way of consequences of our proposal that it be an equiaffine metric, we can but skim only the most obvious ones here. Our discussions are also frustratingly broad-stroked in nature, because we are pulling a bottom block from a massive Jenga and the whole thing has to be rebuilt before we can get to an altitude where definitive and precise experimental tests can be proposed. Even on the mathematical front, the knowledge of equiaffine hypersurfaces (especially those with indefinite $\mathbb{\Pi}_{ab}$) is somewhat limited, due in part to the added difficulties resulting from the raised order of osculating approximates and thus of the differential equations. 
Nevertheless, we think the glimpse of possibilities is sufficiently interesting that we wish to share with the wider community, to solicit interest for further forays.

\acknowledgements
This work is supported by the National Natural Science Foundation of China grants 11503003 and 11633001, the Interdiscipline Research Funds of Beijing Normal University, and the Strategic Priority Research Program of the Chinese Academy of Sciences Grant No. XDB23000000.

\bibliography{SpeedLimit.bbl}

\appendix

\section{Equiaffine geometry} \label{ap:Equiaffine}
Equiaffine differential geometry studies invariants under active global equiaffine transformations of the ambient (specializing to the dimension we are interested in)
\bea
\text{ASL}(5,\mathbb{R}) \equiv \text{SL}(5,\mathbb{R}) \ltimes \mathbb{R}^{5}\,,
\eea
i.e., special linear transformations and translations inside the ambient affine space for locations, accompanied by only the special linear normal subgroup actions on the tensors residing over the tangent and cotangent vector spaces. Quantities and concepts that, when computed and assessed for the post-transformation hypersurface, match the transformation of the original counterparts, are considered invariant. Equiaffine transformations preserve only parallelism, partition ratio of three points, as well as parallelepiped volume and orientation. They thus deform more strongly than rigid rotations of the ordinary Euclidean hypersurface theories, altering in particular orthogonality relations as defined by any ambient metric, which then cannot be used to pick out an equiaffine invariant normal to the hypersurface. 
More precisely, let $\hat{\iota}^{\alpha} = \text{L}_{\beta}{}^{\alpha} \iota^{\beta} + b^{\alpha}$ be an $\text{ASL}(5,\mathbb{R})$ transformation on the immersion function $\iota^{\alpha}$ of a hypersurface (Greek indices run over five dimensions, and Latin four), with $\text{L}_{\beta}{}^{\alpha} \in \text{SL}(5,\mathbb{R})$, then $\hat{\mu}^{\alpha}(\hat{\iota}_q)\neq \text{L}_{\beta}{}^{\alpha}\mu^{\beta}(\iota_q)$, where $\mu^{\alpha}$ ($\hat{\mu}^{\alpha}$) denotes the normal in the sense of being orthogonal to the immersed hypersurface traced out by $\iota$ ($\hat{\iota}$), and $q$ denotes an arbitrary point on the abstract hypersurface (domain of the immersion functions). An alternative approach is therefore needed to arrive at an unique equiaffine invariant normal. 

One begins with a generic transverse vector field $\nu^{\alpha}$. To get to $\mathbb{\Pi}_{ab}$ associated with it, one computes firstly a bilinear form by taking the derivative of a conormal. A conormal $n_{\alpha}$ is an ambient one form or covector, which is $\propto \varphi_{,\alpha}$ if the hypersurface is defined as a level surface of $\varphi$ in a local neighbourhood. The orientation of $n_{\alpha}$ is fixed, but we cannot map it into a normal direction since the natural duality between vectors and covectors is precisely the ambient metric that we don't have. Nevertheless, its directional certainty is sufficient to ensure its contraction (no metrics involved, just covectors acting on vectors as linear functions) with any tangent vector to always vanish, so an ambient derivative along any tangential direction on such a contraction yields zero. On the other hand, application of the Leibniz's rule splits such a derivative into two terms, one with a derivative on the $n_{\alpha}$ half of the contraction which is just that bilinear form we prepared, while the other has the derivative acting on the tangent vector half, with the component along $\nu^{\alpha}$ of the outcome given by the second fundamental form $\mathbb{\Pi}_{ab}$ we seek (i.e., this $\mathbb{\Pi}_{ab}$ is defined as the coefficient in the Gauss formula). Equating the two terms to zero then allows us to extract $\mathbb{\Pi}_{ab}$ from the bilinear form, simply through dividing it by $-n_{\alpha} \nu^{\alpha}$ (which is nonvanishing since $\nu^{\alpha}$ is required to be transverse). 

There are various freedoms not yet fixed in this procedure: while the orientation of $n_{\alpha}$ is certain, we cannot normalize its scaling due to the lack of an ambient metric; both orientation and scaling of $\nu^{\alpha}$ are also free. In order to further pin down a unique equiaffine invariant $\nu^{\alpha}$ with a unique corresponding $\mathbb{\Pi}_{ab}$, we then impose the following conditions to gradually narrow down the choices:

\begin{enumerate}
\item 
An obvious step one can take to partially remove the arbitrariness is by synchronizing $n_{\alpha}$ and $\nu^{\alpha}$ with the condition
\bea \label{eq:Synch}
n_{\alpha} \nu^{\alpha} = 1\,,
\eea
which, for a fixed $n_{\alpha}$, can be seen as a normalization for $\nu^{\alpha}$, but is unfortunately not sufficient to determine $\nu^{\alpha}$ completely, since there isn't a preferred orientation for it. We thus want to supplement and enhance this normalization condition. We do so by taking inspirations from the Euclidean theory, where fixing the amplitude of a normal has the consequence that only tangential components are present in the Weingarten equation governing the derivative of said normal along the hypersurface. We impose the same condition on $\nu^{\alpha}$, and call the transverse fields satisfying this condition relative normals. This strategy works, yielding a unique $\nu^{\alpha}$ to any given $n_{\alpha}$, and allows for 
\bea\label{eq:Shape}
\mathbb{S}^a{}_{b}\equiv -\nu^{a}{}_{,b}
\eea
to serve as the shape operator, in a fashion analogous to the ordinary Euclidean theory. 

Furthermore, it is equivalent \cite{Nom1} to demanding that the hypersurface volume form $\tilde{\omega}$ induced by $\nu^{\alpha}$ should be parallelly transported by the intrinsic covariant derivative $\tilde{\nabla}_a$, induced from the flat ambient connection by borrowing the expression for the Gauss formula from the ordinary Euclidean theory. The volume $\tilde{\omega}$ is defined with determinants (top differential forms), by eliminating, through contracting with $\nu^{\alpha}$, a conormal contribution from within the ambient volume form. Eq.~\eqref{eq:Synch} ensures that this conormal is in fact $n_{\alpha}$.  

\item 
There is only then the scaling freedom in $n_{\alpha}$, or equivalently a conformal freedom in $\mathbb{\Pi}_{ab}$ \cite{AffineGeo}, that still needs to be fixed. Because $\tilde{\omega}$ is obtained by factoring out $n_{\alpha}$ from the ambient volume form, a condition on $\tilde{\omega}$ would 
be quite effective. For the equiaffine case (there are other conditions leading to sibling theories, like centroaffine or Euclidean-affine, whose unique normals are invariant under different subsets of the general linear plus translation transformations), we impose the apolarity condition, requiring $\tilde{\omega}$ to agree with the intrinsic pseudo-Riemannian volume form ${}_{\mathbb{\Pi}}\omega \propto \sqrt{|\det(\mathbb{\Pi}_{ab})|}$. This provides some assurances that ${}_{\mathbb{\Pi}}\omega$, even though defined through an indefinite $\mathbb{\Pi}_{ab}$ that can output negative distances, would not end up straying too far from how one expects a volume to behave. 
There is a unique relative normal satisfying this condition, called the equiaffine (or Blaschke) normal, given by $\nu^{\alpha}={}_{\mathbb{\Pi}}\Box \iota^{\alpha}/4$. Roughly, it finds a central direction around which the hypersurface locally looks as symmetric as possible -- see \cite{BLA1,LEI} for more rigorous descriptions. As equiaffine transformations sheer the hypersurface shape, the central direction tilts concomitantly, allowing this particular relative normal to be an equiaffine invariant. 

Note, this is generally a different fixing than the Euclidean normal, wherever an ambient metric is provided, so one has to be careful about borrowing intuitions from the Euclidean theory. Nevertheless, the Euclidean normal is also a legitimate relative normal, and thus the $\mathbb{\Pi}_{ab}$s corresponding to the Euclidean and equiaffine normals are conformally related, and many important properties, such as rank and signature (modulo interchanging the positive and negative slots), are shared between them. However, it is the first fundamental form and not the second that's used as the intrinsic metric for the ordinary Euclidean differential geometry, so that theory is different from the Euclidean normal-fixed version of affine geometry (viz.~Euclidean-affine), and its intrinsic pseudo-Riemannian (recall Footnote \ref{fn:naming}) counterpart theory is not conformally related to the intrinsic partner of the equiaffine theory.

To implement the apolarity condition, we need to bring in the totally symmetric cubic form $\mathbb{F}_{abc}$ that evaluates the breakdown of metricity, when that ambient-induced covariant derivative $\tilde{\nabla}_a$ of the last enumeration point is used on the candidate equiaffine metric $\mathbb{\Pi}_{ab}$, i.e., 
\bea \label{eq:FDown}
\mathbb{F}_{abc}=-\frac{1}{2}\tilde{\nabla}_a\mathbb{\Pi}_{bc}\,. 
\eea
Raising indices using $\mathbb{\Pi}^{ab}$ defined through $\mathbb{\Pi}^{ab}\mathbb{\Pi}_{bc} = \delta^{a}_{c}$ from here on and throughout the paper, we obtain $\mathbb{F}_{ab}{}^c$ that measures the difference between the ambient-induced connection (associated with $\tilde{\nabla}_a$), formally written in component form as $\tilde{\Gamma}_{ab}^{c}$, and the Christoffel symbol ${}_{\mathbb{\Pi}}\Gamma_{ab}^{c}$ of $\mathbb{\Pi}_{ab}$, viz., 
\bea \label{eq:CubicFormDef}
\mathbb{F}_{ab}{}^{c} \equiv  \tilde{\Gamma}_{ab}^{c} -{}_{\mathbb{\Pi}}{\Gamma}_{ab}^{c}\,.
\eea
The apolarity condition is then imposed as the algebraic relation 
\begin{align} \label{eq:apolarity}
0=&\mathbb{\Pi}^{bc}\mathbb{F}_{abc} = \mathbb{F}_{ab}{}^{b} 
= \tilde{\Gamma}_{ab}^{b} -{}_{\mathbb{\Pi}}{\Gamma}_{ab}^{b}  \notag \\
=& \left(\ln \tilde{\omega}\right)_{,a} - \left(\ln {}_{\mathbb{\Pi}}{\omega}\right)_{,a} = \left(\ln \frac{\tilde{\omega}}{{}_{\mathbb{\Pi}}{\omega}}\right)_{,a}\,,
\end{align}
or the ability to propagate an ${}_{\mathbb{\Pi}}\omega = \tilde{\omega}$ initial fixing from one location to all across the (assuming connected) hypersurface.

\end{enumerate}

\end{document}